\documentclass[11pt]{article}
\usepackage {amsmath}
\usepackage[dvips]{graphicx}
\usepackage{feynmp}
\usepackage{amssymb}
\usepackage{cite}
\usepackage{appendix}
\newcommand{\be}{\begin{equation}}
\newcommand{\ee}{\end{equation}}
\newcommand{\bear}{\begin{eqnarray}}

\newcommand{\eear}{\end{eqnarray}}

\newcommand{\vev}[1]{\left\langle #1\right\rangle}

\newcommand{\TeV}{\; \mathrm{TeV}} 
\newcommand{\lapproxeq}{\lower .7ex\hbox{$\;\stackrel{\textstyle  
<}{\sim}\;$}} 
\newcommand{\gapproxeq}{\lower .7ex\hbox{$\;\stackrel{\textstyle  
>}{\sim}\;$}} 
\newcommand{\stackdown}[2]{\lower 1.4ex\hbox{$\;\stackrel{\textstyle{#1}}  
{\scriptstyle{#2}}\;$}}
\newcommand{\beq}{\begin{equation}} 
\newcommand{\eeq}{\end{equation}} 

\newcommand{\ba}{\begin{eqnarray}}
\newcommand{\ea}{\end{eqnarray}}

\newcommand{\bea}{\begin{eqnarray}}
\newcommand{\eea}{\end{eqnarray}}

\makeatletter 
\def\slash{\@ifnextchar[{\fmsl@sh}{\fmsl@sh[0mu]}} 
\def\fmsl@sh[#1]#2{%
  \mathchoice 
    {\@fmsl@sh\displaystyle{#1}{#2}}%
    {\@fmsl@sh\textstyle{#1}{#2}}%
    {\@fmsl@sh\scriptstyle{#1}{#2}}%
    {\@fmsl@sh\scriptscriptstyle{#1}{#2}}} 
\def\@fmsl@sh#1#2#3{\m@th\ooalign{$\hfil#1\mkern#2/\hfil$\crcr$#1#3$}} 
\makeatother 
\begin{document}
\newpage 
\baselineskip=18pt 
\begin{titlepage}  
\begin{flushright} 
\end{flushright} 
\begin{flushright} 
\parbox{4.6cm}{UA-NPPS/BSM-2-13 }
\end{flushright}
\vspace*{5mm} 
\begin{center} 
{\large{\textbf {
Six-dimensional origin of gravity mediated brane to brane supersymmetry breaking
}}}\\
\vspace{14mm} 
{\bf G. A.~\ Diamandis}, \, {\bf B. C.~\ Georgalas}, \, 
{\bf P.~\ Kouroumalou } and \\ {\bf A. B.~\ Lahanas}
{\footnote{email alahanas@phys.uoa.gr}}
\vspace*{6mm} \\
  {\it University of Athens, \\Physics Department,  
Nuclear and Particle Physics Section,\\  
GR--15771  Athens, Greece}
\end{center} 
%
\begin{abstract}
Four dimensional supergravities may be the right framework to describe particle physics at low energies. Its connection to the underlying string theory can be implemented through higher dimensional  supergravities  which bear special characteristics. Their reduction to four dimensions breaks supersymmetry whose magnitude depends both on the compactifying manifold and the mechanism that generates the breaking. In particular compactifications, notably on a $S_1/Z_2$ orbifold, the breaking of supersymmetry occuring on a hidden brane,  residing at one end of  $S_1/Z_2$, is communicated to the visible brane which lies at the other end, via gravitational interactions propagating in the bulk. This scenario has been exemplified in the framework of the $N=2$, $D=5$ supergravity. In this note, motivated by the recent developments in the field, related to the six-dimensional description of the supergravity theory, we study the $N=2$, $D=5$ supergravity theory as originating from a $D=6$ supergravity which, in addition to the gravity, includes a number of tensor multiplets. This reduces to $N=1$, $D=4$ supergravity in a two step manner, first by Kaluza-Klein reduction followed by a  $S_1/Z_2$ orbifold compactification. The resulting  theory has striking similarities with the one that follows from the single standalone $N=2$, $D=5$ supergravity, with no reference to the underlying higher dimensional $D=6$  supergravity,  and a structure that makes  the supersymmetry breaking mechanisms studied in the past easily incorporated in higher dimensional schemes. 
\end{abstract} 
\end{titlepage} 
\newpage 
\baselineskip=18pt 

\section{Introduction}

 The study of supergravities in five and six dimensions has been revived in  recent years for various reasons. In particular, six dimensional supergravity theories naturally arise as realizations of F-theory \cite{Vafa1} while five-dimensional theories may emerge as limits of M-theory \cite{horava, recent, fived}. Also the duality between F and M theory is better studied via the reduction of six-dimensional to five-dimensional supergravity \cite{Ferrara, grimm1, grimm2}.  Besides it has been long known  that five-dimensional supergravity compactified on $S_1/Z_2$ may arise as the effective theory of stongly coupled heterotic string theory \cite{ovrut}. The latter has stimulated the interest towards  studying the  phenomenological and cosmological \cite{roy, pal} consequences of supersymmetric brane world models. In the framework of these models, the  $S_1/Z_2$
orbifold compactification determines effectively two spatially three-dimensional branes  one of which is identified as the visible world and the other is the invisible brane or the so called hidden brane. An important issue, which has been extensively studied in this context, is the origin of supersymmetry breaking. Towards that goal there are several mechanisms proposed in literature \cite{kane, hiddensec, large2,  peskin, bagger1, nilles,  lalak,  radion1, meissner, okada, gates,  anomalymed1, anomalymed2, riotto, rattazzi,scrucca, gregoire, tricher, SchmidtHoberg:2005yy, peggy1, peggy2, peggy3, McGarrie:2010kh, McGarrie:2010yk, Bouchart:2011va} and a popular scenario is the one in which the supersymmetry breaking takes place  at the hidden brane and this breaking is radiatively communicated through the bulk  to the visible world. In this context the five-dimensional theory considered is mainly pure supergravity and thus the interactions propagating in the five-dimensional space is gravity and the radion multiplet connected to the geometry of the fifth dimension. The latter acts as a messenger of the supersymmetry breaking, along with the five-dimensional gravity multiplet,  via loop corrections,  from the hidden to the visible brane where the observable fields live. A striking feature of this mechanism is the fact that the resulting supersymmetry breaking scale turns out to be finite, due to cancellations between  bosonic and fermionic contributions occurring in loops, and also to the fact that the size of the orbifold is non-vanishing setting the separation of the two branes and the scale of supersymmetry (SUSY) breaking. 

The aforementioned  scheme was exemplified in  previous works \cite{peggy2, peggy3}, where observable fields were assumed to live on the visible brane interacting with the bulk gravity which mediates supersymmetry breaking which takes place at the hidden brane. It would be interesting to see if this mechanism can be generalized in such a way that is embedded in the framework of  a higher dimensional supergravity theory and if so what are the consequences for the supersymmetry breaking scale and its implications for phenomenology. 
In this note we undertake this problem and investigate  the possibility of the transmission of the supersymmetry breaking in models originated from the six-dimensional (1, 0) supergravity \cite{Nishino1, romans, Nishino2, Schwarz1, Seiberg1,  Nishino3, riccioni1, riccioni2, Avramis1, Avramis2, peggy2013}. In particular we take up the point that  the N=2 five-dimensional supergravity \cite{fivesugra, gunaydin} is originating by Kaluza-Klein compactification of a simple (1, 0) six-dimensional  supergravity with tensor multiplets present. We find that the $S_1/Z_2$ orbifolding shares the basic features which permit the transmission of the  supersymmetry breaking from the hidden   to the visible brane via fields that live  on the branes and are propagating in the bulk. However a basic difference emerges from the purely five-dimensional considerations, pertinent to the existence of many chiral multiplets involved, which are unavoidable when we compactify  from higher dimensions. 
Their projections to the branes is described by a K\"{a}hler function reminiscent of the no-scale type \cite{noscale}. Due to this structure, the introduction of a superpotential on the branes leads to positive definite scalar potential, as in the pure five-dimensional case, while  considering a constant superpotential, in order to generate the supersymmetry breaking on the hidden brane, gives no potential on it. These  basic ingredients for the transmission of supersymmetry breaking  are preserved in the 6-dimensional considerations and the successful predictions of the five-dimensional models remain in principle intact.

\section{$\, N=2, D=5$ supergravity and its $D=4$ reduction}

In an earlier paper \cite{peggy2} we studied the reduction of the five-dimensional supergravity on $\, S_1/Z_2 \,$ orbifold the ends of which 
may be conceived as two branes. We assumed that only gravity propagates in the fifth dimension and the observable matter is located on one of the two branes although other more complicated  options are available. The Lagrangian is of the form $\, {\cal{L}} =  {\cal{L}}_5 + {\cal{L}}_b \,$ where $\, {\cal{L}}_5  \,$ is the Lagrangian part describing $\, N=2 , D=5 \,$  supergravity and $\, {\cal{L}}_b  \,$
the  part including the interactions of all fields living on the brane among themselves as well as their  interactions  with the projections of the bulk fields on the brane. 
$\, {\cal{L}}_b  \,$ has the structure of a four-dimensional supergravity which includes except the observable  chiral and vector multiplets the radion  multiplet.  In particular was  found that the Lagrangian part $\, {\cal{L}}_b  \,$ can be described by a K\"{a}hler  function 
\be
{\cal{F}} \, = \, -3 \, ln \, \frac{T+T^*}{\sqrt{2}} \, + \, \frac{\sqrt{2}}{T+T^*} \, \left( \, \delta(x_5) \,  K_V( \phi_V, \phi^*_V \,) \, + \,
\delta(x_5 - \pi \, R ) \,  K_H( \phi_H, \phi^*_H \,) \,  \right)
\label{branepast}
\ee
where $\, K_i( \phi_i, \phi^*_i) $ are  the K\"{a}hler  functions, in general,  of the fields, $\, i=V$, that are located on the visible brane at $\, x_5=0$, and those that are located on the hidden brane,  $\, i = H \, $, which lies at $\, x_5 = \pi \, R \, $ . In (\ref{branepast}) the scalar $\, T$ belongs to the radion multiplet which with a proper normalization has components $\, T, \,  \chi_T$ given by 
\be
T \, =\, \frac{1}{\sqrt{2}} \, \left( \, e_{\, 5}^{\, \dot{5}} \, + i \, \sqrt{ \frac{2}{3} } \, A_{\,5}^{\, 0} \, \right) \quad , \quad 
\chi_T \, = \, - \psi_{\, 5}^{\,2}
\label{radion}
\ee
In it $\, e_{\, 5}^{\, \dot{5}} $ is the vielbein component associated with  the fifth dimension, $\,A_{\,5}  $ is the fifth component of the gravi-photon and similarly $\,\psi_{\, 5}^{\,2}  $ is the fifth component of the second of the gravitino fields ( for details see \cite{peggy2} ). 
Besides,  superpotential couplings, which are descibed by proper superpotentials $\,W_V( \phi_V \,)  $ and $\,W_H( \phi_H \,)  $, have to be added in order to complete the picture.  
These results hold to order $\, k_5^2 $, where $\, k_5^2 $  is the five-dimensional Newton's constant, and for their derivation we worked in the on-shell five-dimensional supergravity using Noether's method \cite{vann}. 

The aforementioned  scheme was employed in order to study the brane to brane mediation of the  supersymmetry breaking mechanism. 
In order to  pave the ground for the discussion that follows we only recall here the salient qualitative features of the mechanism deployed in previous works \cite{peggy2,peggy3}, where the above mechanism was employed. The hidden and the visible branes, located at the end points of $\, S_1/Z_2 \,$,  communicate through gravitational interactions which, we assume,  are the only forces that propagate in the bulk and the only ones that can convey the message of supersymmetry breaking from the hidden brane to the observable fields that are located on the visible brane. The orbifolding breaks one of the two supersymmetries of the $N=2$ supergravity, while the remaining local supersymmetry is broken on the hidden brane, giving mass to the gravitino, by a properly chosen superpotential $\, W_H$, and this breaking is transmitted to the visible fields. This idea has been exemplified in a simple model \cite{peggy3} and it has been shown that finite  soft supersymmetry breaking terms can be induced which depend on the gravitino mass $\, m_{3/2}$, and the size of the orbifold $\, S_1/Z_2 \,$. 
Therefore the seeds of supersymmetry breaking reside on the hidden brane, where a superpotential breaks local supersymmetry, and it is essential that the brane-bulk field  interactions are fully known and in particular those that are relevant for the transmission of supersymmetry breaking from one brane to the other. 
 A convenient option for the hidden superpotential is to take it to be a constant $\, W_H = c \,$ \cite{riotto,bagger1}  whose value is directly related to the gravitino mass by 
$\, m_{3/2} \, = \, k_5^2 \, c / \pi R \, $, where  $\, \pi R $ sets the size of $\, S_1/Z_2 \,$ 
{\footnote{
$m_{3/2} \,$ is actually the mass of the lowest gravitino mode, in the approximation $\, k_5^2 \, c < 1 \,$, or same $m_{3/2} \, \lesssim  \, R^{\, -1}$.
\label{footot}
}}
. 
Therefore the gravitino mass 
$\, m_{3/2}  \, $  can be traded for  the  parameter  $\, c$ which is set by the constant superpotential. 
The supersymmetry breaking is transmitted to the observable fields on the visible brane via gravitational interactions through loops including the graviton and gravitino fields as well as the fields of the radion multiplet. We remark that the role of the radion multiplet is essential in this approach since its presence yields contributions that can make the soft masses squared of the squarks and sleptons be non-tachyonic.  
In \cite{peggy2, peggy3}  we exemplify the above mechanism in a simple prototype and showed that finite soft supersymmetry breaking parameters are indeed generated by loops, which involve the gravitino and the fermion of the radion multiplet, and their magnitudes can be determined. 

After this  outline of the salient features of the five-dimensional supergravity and its role for the brane to brane mediation of the supesymmetry breaking  it is essential to explore whether this scheme can be incorporated, if it does at all, in the framework of higher dimensional supergravities as discussed in the introduction. This is important if we believe that the low energy supersymmetry has its origin in a more fundamental theory, like string theory for instance , which includes gravity and has supersymmetric structure. 

In the following section we shall embark on first discussing the six-dimensional supergravity, which is the simplest higher dimensional scheme in which the five-dimensional supergravity can be embedded and its reduction following the Kaluza - Klein ansatz.

\section{From six to five-dimensional supergravity via  Kaluza-Klein reduction}

Our starting point is  the simplest six-dimensional supergravity which, except the gravity multiplet, we assume it includes additional  tensor multiplets. 
The role of tensor multiplets is essential for the cancellation of anomalies in higher supergravities, as well as the presence of hypermultiplets. 
In this work we do not proceed to a full investigation of the problem and the construction of an anomaly-free model. However we allow for the existence of tensor multiplets to preserve the qualitative features of an anomaly-free theory \cite{Schwarz1, Avramis1, Avramis2}. Some models demand  more than one tensor multiplets to guarantee absence of anomalies, however we start with the least content, allowing for one tensor multiplet, and later we  generalize the result for more multiplets
{\footnote{
In our considerations for simplicity we assume that only gravity propagates in the bulk. Cases where in addition to gravity gauge fields  propagate in the bulk are not considered in this note. 
}}.
Therefore in this consideration the spectrum  consists of the supergravity multiplet
\be
\hat{e}_{\hat{M}}^{\hat{m}}, \,\, \hat{\Psi }_{\hat{M}}, \,\, \hat{B}_{\hat{M} \hat{N}}^1
\ee
and a tensor multiplet
\be
\hat{B}_{\hat{M} \hat{N}}^2, \,\,\varphi, \,\, \hat{X }
\ee
where 
$\hat{B}_{\hat{M} \hat{N}}^{1,2}$ have  field strengths 
\be
 \hat{H}_{\hat{M} \hat{N} \hat{P}}^{1,2} \,=\, 3\partial_{[\hat{M}} \hat{B}^{1,2}_{\hat{N} \hat{P}]}
 \nonumber
\ee
obeying the duality conditions
\be
G_{rs} \hat{H}^{s\, \hat{M} \hat{N} \hat{P}} \,=\, \eta _{rs} E^{\hat{M} \hat{N} \hat{P} \hat{K} \hat{\Lambda } \hat{\Sigma }} \hat{H}^s _{\hat{K} \hat{\Lambda } \hat{\Sigma }} \quad .
\nonumber
\ee
In the equations above  $r, s = 1, 2 \,$, with  $\,\eta _{rs} = diag(1, -1)  \,$ and ( see for instance\cite{romans} )
\be
G_{11} = G_{22} = \frac{1}{2} \left( e^{2\varphi } + e^{-2\varphi }  \right) , \quad \quad G_{12} = \frac{1}{2} \left( e^{2\varphi } - e^{-2\varphi }  \right)
\ee
The fields $ \hat{\Psi }_{\hat{M}},  \hat{X }$ are left(right)-handed symplectic Majorana spinors.

The bilinear terms of the relevant six-dimensional Lagrangian are 
\bea
\hat{e}^{-1} \mathcal{L}^{(6)} \,&=&\, \frac{1}{2} R^{(6)} (\hat{G}) \,-\, \frac{1}{4} \partial _{\hat{M}} \varphi  \partial ^{\hat{M}} \varphi \,-\, 
\frac{1}{12} G_{rs} \hat{H}^{r}_{\hat{M} \hat{N} \hat{P}} \hat{H}^{s \,\hat{M} \hat{N} \hat{P}} \nonumber \\
&{}& \nonumber \\
&& + \, \frac{i}{2} \; \bar{\Psi }_{\hat{M}} \Gamma ^{\hat{M} \hat{N} \hat{P}} D_{\hat{N}} \Psi _{\hat{P}} \,- \, \frac{i}{2} \;  \bar{X}\Gamma ^{\hat{M}} D_{\hat{M}} X \quad .
\label{sixdl}
\eea
In this, $\, \Gamma ^{\hat{M}}  $ are six-dimensional gamma matrices and  $\, \Gamma ^{\hat{M} \hat{N} .... \hat{P}} $    denote their properly weighted  antisymmetric products. 
Note that the tensor field kinetic terms for the case under consideration case may be put in a diagonal form  
\[
- \frac{1}{24} \, e^{ 2 \varphi } \hat{G}^{(1), \, \hat{M} \hat{N} \hat{P}} \hat{G}_{\hat{M} \hat{N} \hat{P}}^{(1)} - \frac{1}{24} \, e^{-2 \varphi } \hat{G}^{(2), \, \hat{M} \hat{N} \hat{P}} 
\hat{G}_{\hat{M} \hat{N} \hat{P}}^{(2)}
\]
by defining 
$ \hat{G}_{\hat{M} \hat{N} \hat{P}}^{(1)} $   and $ \hat{G}_{\hat{M} \hat{N} \hat{P}}^{(2)} $ as 
\[
\hat{G}_{\hat{M} \hat{N} \hat{P}}^{(1)} = \hat{H}_{\hat{M} \hat{N} \hat{P}}^1 + \hat{H}_{\hat{M} \hat{N} \hat{P}}^2 \quad \quad , \quad \quad \hat{G}_{\hat{M} \hat{N} \hat{P}}^{(2)} = \hat{H}_{\hat{M} \hat{N} \hat{P}}^1 - \hat{H}_{\hat{M} \hat{N} \hat{P}}^2 \quad .
\]

The supersymmetry transformations of the bosonic components  are given by 
\begin{eqnarray}
\delta \hat{e}_{\hat{M}}^{\hat{m}} \hspace*{3mm} &=&  i \, \bar{E} \, \Gamma ^{\hat{M}} \Psi_{\hat{M}} \nonumber \\
\delta \, \varphi \hspace*{5mm} &=&  \bar{E} X \nonumber \\
\delta \hat{B}_{\hat{M} \hat{N}} &=& - \frac{i}{2} \, e^{-\varphi } \left( \bar{E}\Gamma _{\hat{M}} \Psi _{\hat{N}} \,-\,
\bar{E}\Gamma _{\hat{N}} \Psi _{\hat{M}} \,- \, i \bar{E}\Gamma _{\hat{M} \hat{N}} X \right) \nonumber \\
\delta \hat{C}_{\hat{M} \hat{N}} &=& -\frac{i}{2} \, e^{\varphi } \left( \bar{E}\Gamma _{\hat{M}} \Psi _{\hat{N}} \,-\,
\bar{E}\Gamma _{\hat{N}} \Psi _{\hat{M}} \,+\, i \bar{E}\Gamma _{\hat{M} \hat{N}} X \right) 
\label{trans}
\end{eqnarray}
where
\[
\hat{B}_{\hat{N} \hat{P}} \,=\, \hat{B}_{\hat{N} \hat{P}}^1 + \hat{B}_{\hat{N} \hat{P}}^2
\]
and
\[
\hat{C}_{\hat{M} \hat{N}} \,=\, \hat{B}_{\hat{N} \hat{P}}^1 - \hat{B}_{\hat{N} \hat{P}}^2 \quad .
\]
In the transformation given by Eq.(\ref{trans}),  the parameter defining the local supersymmetry transformations   $E$ is a left-handed symplectic Majorana spinor.

The standard Kaluza-Klein ansatz reduces the action to five dimensions. The field modes  are considered not to depend on the sixth dimension and in particular the metric is decomposed in the following manner
\begin{eqnarray}
\hat{G}_{\hat{M} \hat{N}} &\mapsto & \left\{\hat{G}_{MN} = G_{MN} + e^{2\sigma }A_{M}A_{N}, \,\,\hat{G}_{M6} = e^{2\sigma }A_{M}, \,\,\hat{G}_{66} = e^{2\sigma } \right\} \nonumber \\
&{}& \nonumber \\
\hat{G}^{\hat{M} \hat{N}} &\mapsto & \left\{\hat{G}^{MN} = G^{MN}, \,\,\hat{G}^{M6} = - A^{M}, \,\,\hat{G}^{66} = e^{-2\sigma }+A^2 \right\} 
\label{redme}
\end{eqnarray}
The above decomposition corresponds to the following decomposition of the "sechsbein"
\begin{eqnarray}
\hat{e}_{\hat{M}}^ {\hat{m}} &\mapsto & \left\{\hat{e}_{M}^{\tilde{m}} = e_{M}^{\tilde{m}}, \,\,\hat{e}_{M}^{\dot{6}} = e^{\sigma }A_{M}, \,\,\hat{e}_{6}^{\tilde{m}} = 0, \,\, {e}_{6}^{\dot{6}} = e^{\sigma } \right\} \nonumber \\
&{}& \nonumber  \\
\hat{e}_{\hat{m}}^ {\hat{M}} &\mapsto & \left\{\hat{e}_{\tilde{m}}^{M} = e_{\tilde{m}}^{M}, \,\,\hat{e}_{\dot{6}}^{M} = 0, \,\,\hat{e}_{\tilde{m}}^{6} = -A_{\tilde{m}}, \,\, {e}_{\dot{6}}^{6} = e^{-\sigma } \right\} 
\end{eqnarray}
This reduction leads to a five-dimensional supergravity \cite{Ferrara, grimm1} with the presence of two vector multiplets furnished by the fields 
\be
A^1_M = B_{M6}, \,\, A^2_M = C_{M6}, \,\, \varphi , \,\, \sigma , \,\, \Psi _6, \,\, X \, .
\nonumber
\ee
In order to bring  the five-dimensional gravitational action to the Einstein-Hilbert form the following  rescaling is necessary
\be
e_{M}^{\tilde{m}} \,=\, e^{-\sigma /3} \tilde{e}_{M}^{\tilde{m}}
\label{wres}
\ee
That done  the bilinear terms of the bosonic fields take on the form
\begin{eqnarray}
\mathcal{L}^{(5)}_{0b} \,&=&\, \sqrt{-\tilde{G}} \,\, \bigg[ \frac{1}{2} \, R^{(5)}(\tilde{G}) - \frac{1}{3} \, \tilde{G}^{MN} \partial _M \sigma  \partial _N \sigma 
- \frac{1}{4} \, \tilde{G}^{MN} \partial _M \varphi  \partial _N \varphi \nonumber \\
&-& \frac{1}{16} \, e^{2\sigma /3} \, F_{MN} F^{MN} - \frac{1}{12} \, 
e^{2\varphi - 4\sigma /3} \, F^{1}_{MN} F^{1\,MN} - \frac{1}{12} 
e^{-2\varphi - 4\sigma /3} \, F^{2}_{MN} F^{2\,MN} \bigg]
\label{fivedl}
\end{eqnarray}
where $F^{1,2}_{MN}$ are  the field strengths of the vector fields $A^{1,2}_M \, $ respectively and $F_{MN}$ the field strength of $\, A_M$ stemming from the reduction of the  metric 
( see Eq. (\ref{redme}) ). 
Notice that the $\sigma$ field kinetic term appear after the rescaling (\ref{wres}). 
Moreover in order to have a canonical kinetic term  and the standard supersymmetry transformation law for the five-dimensional gravitino field
we have to  make the following redefinition 
\be
\tilde{\Psi }_M \,=\, e^{\sigma /6} \Psi _M - \frac{1}{3} (e_6^{\dot{6}})^{-1} \Gamma _{\tilde{n}} \Gamma ^{\dot{6}} \tilde{\Psi }_6 \tilde{e}_{M}^{\tilde{n}}
\ee
where $\tilde{\Psi }_M \,=\, e^{-\sigma /6} \Psi _M$. Rescaling  the parameter of the supersymmetry transformation as 
\be
E \,=\, e^{-\sigma /6} \tilde{E}
\nonumber
\ee
we recast the ordinary transformation law for the "f\"{u}nfbein", i.e.
\be
\delta \tilde{e}_{M}^{\tilde{m}} \hspace*{3mm} =   i \, \tilde{E} \, \Gamma ^{\tilde{m}} \tilde{\Psi}_{M}
\ee

\section{$S_1/Z_2$ orbifold and coupling of brane fields}

The orbifolding is implemented by decomposing the five-dimensional metric and  assigning proper $Z_2$-parities to the fields involved. In particular for the  metric we decompose as:
\begin{eqnarray}
\tilde{G}_{M N} &\mapsto & \left\{\tilde{G}_{\mu \nu } = \tilde{g}_{\mu \nu } + e^{2 \tau  }B_{\mu }B_{\nu }, \,\,
\tilde{G}_{\mu 5} = e^{2 \tau  }B_{\mu }, \,\,\tilde{G}_{55} = e^{2 \tau  } \right\} \nonumber  \\
&{}& \nonumber \\
\tilde{G}^{MN} &\mapsto & \left\{\tilde{G}^{\mu \nu } = \tilde{g}^{\mu \nu }, \,\,\tilde{G}^{\mu 5} = - B^{\mu }, \,\,
\tilde{G}^{55} = e^{-2 \tau  }+B^2 \right\}.
\end{eqnarray}
The above decomposition corresponds to the following decomposition of the "f\"{u}nfbein"
\begin{eqnarray}
\tilde{e}_{M}^ {\tilde{m}} &\mapsto & \left\{\tilde{e}_{\mu }^{\dot{m}} = \tilde{e}_{\mu }^{\dot{m}}, \,\,\tilde{e}_{\mu }^{\dot{5}} = e^{\tau  }B_{\mu }, \,\,\tilde{e}_{5}^{\dot{m}} = 0, \,\, \tilde{e}_{5}^{\dot{5}} = e^{\tau  } \right\} \nonumber \\
&{}& \nonumber  \\
\tilde{e}_{\tilde{m}}^ {M} &\mapsto & \left\{\tilde{e}_{\dot{m}}^{\mu } = \tilde{e}_{\dot{m}}^{\mu }, \,\,\tilde{e}_{\dot{5}}^{\mu } = 0, \,\,\tilde{e}_{\dot{m}}^{5} = -B_{\dot{m}}, \,\, \tilde{e}_{\dot{5}}^{5} = e^{-\tau  } \right\} 
\end{eqnarray}
The parity assignments, including these for the "graviphoton" $A_M$, go as follows 
\bea
&& \text{Even fields : } \nonumber \\
&& \text{bosons} \, \quad  \longmapsto \quad \{ \, \tilde{e}_{\mu }^{\dot{m}}, \,\tilde{e}_{\dot{5}}^{5},  \,\sigma , \, \varphi , \, A_5, \, A_{5}^1, \, A_{5}^{2} \, \}
\nonumber \\
&& \text{fermions} \, \; \longmapsto \quad \{ \, \tilde{\psi} _{\mu }^1, \, \tilde{\psi} _{6 }^2, \, \tilde{\psi} _{5 }^2, \, \tilde{\chi} ^2 \, \}
\nonumber
\eea
and
\bea
&& \text{Odd fields : } \nonumber \\
&& \text{bosons} \, \quad  \longmapsto \quad \{ B_{\mu } , \, A_{\mu} , \, A_{\mu }^1, \, A_{\mu }^{2} \, \} \nonumber \\
&& \text{fermions} \, \; \longmapsto \quad \{ \, \tilde{\psi} _{\mu }^2, \, \tilde{\psi} _{6 }^1, \, \tilde{\psi} _{5 }^1, \, \tilde{\chi} ^1 \, \}
\nonumber
\eea
The two supersymmetries, in the chiral  $D=6$  $( 1, 0 )$ supergravity, have parameters the eight-component left-handed spinors
\be
\tilde{E} ^{1,2} \,=\, \left(   \begin{array}{c} {\varepsilon }^{1,2}  \\  0   \end{array} \right)        \quad . 
\nonumber
\ee
These are decomposed in five-dimensions in terms of two-component symplectic Majorana spinors in the following way, 
\bea
\varepsilon  ^1 \,=\, \left(   \begin{array}{c} \varepsilon  \\  \zeta    \end{array} \right), \quad 
\varepsilon  ^2 \,=\, \left(   \begin{array}{c} \zeta  \\  -\varepsilon   \end{array} \right), 
\quad \bar{\varepsilon }_1 \,=\, (\zeta  , \,\, \bar{\varepsilon}), \quad \bar{\varepsilon }_2 \,=\, (-\varepsilon  , \,\, \bar{\zeta  }).
\eea
$ {\varepsilon } \,$ is $Z_2$-even and $\zeta$ is $Z_2$-odd so that supersymmetric transformations of the surviving $N=1$ supersymmetry on the branes, after orbifolding, are implemented by $ {\varepsilon } \,$.

The coupling of the  brane chiral multiplets in this model is obtained by Noether's  method \cite{vann} which is more  suitable for the case 
of on-shell supergravities. For details we point the reader to literature \cite{rattazzi,  peggy2,  belyaev,  benakli}.
We first consider the  restriction of the bosonic kinetic terms of the even fields on the branes, introduced by the orbifolding, which are given by 
the following terms,
\begin{eqnarray}
\mathcal{L}^{(5)}_{0 \, b} \,&=&\, \sqrt{-\tilde{g}} e^{\tau }\,\, \bigg[ \frac{1}{2}R^{(4)}(\tilde{g}) - \frac{1}{6} \tilde{g}^{\mu \nu } \partial _\mu  \sigma  \partial _\nu  \sigma 
- \frac{1}{4} \tilde{g}^{\mu \nu } \partial _\mu  \varphi  \partial _\nu  \varphi -\frac{1}{8} e^{2\sigma /3} e^{-2\tau } \tilde{g}^{\mu \nu } \partial _\mu A_5  \partial _\nu  A_5  \nonumber \\
&-&  \, \frac{1}{6} e^{2 \varphi - 4\sigma /3} e^{-2\tau } \tilde{g}^{\mu \nu } \partial _\mu A_5^1  \partial _\nu  A_5^1 
- \frac{1}{6} e^{-2 \varphi - 4\sigma /3} e^{-2\tau } \tilde{g}^{\mu \nu } \partial _\mu A_5^2  \partial _\nu  A_5^2
 \bigg].
\label{kine}
\end{eqnarray}
A rescaling  of the four dimensional metric given by  
\be
\tilde{g}_{\mu \nu } \,=\, e^{-\tau } g_{\mu \nu }
\nonumber
\ee
brings the terms given by Eq. (\ref{kine}) to the following form 
\begin{eqnarray}
\mathcal{L}^{(5)}_{0 \, b} \, = &&
\sqrt{-g} \,  \bigg[ \; \; {\frac{1}{2}}\, R^{(4)}(g) - \frac{1}{6} g^{\mu \nu } \partial _\mu  \sigma  \partial _\nu  \sigma 
- \frac{1}{4} {g}^{\mu \nu } \partial _\mu  \varphi  \partial _\nu  \varphi
-\frac{3}{4} g^{\mu \nu } \partial _\mu  \tau  \partial _\nu  \tau  \hspace*{6mm}  \nonumber \\
&& \quad - \, \frac{1}{8} e^{2\sigma /3 } e^{-2\tau} {g}^{\mu \nu } \partial _\mu A_5 \, \partial _\nu  A_5  - 
\frac{1}{6} e^{2 \varphi - 4\sigma /3} e^{-2\tau } {g}^{\mu \nu } \partial _\mu A_5^1 \, \partial_\nu  A_5^1 \nonumber \\
&& \quad - \frac{1}{6} e^{-2 \varphi - 4\sigma /3} e^{-2\tau } {g}^{\mu \nu } \partial _\mu A_5^2 \, \partial _\nu  A_5^2 \; \;
 \bigg]
\end{eqnarray}
It is interesting to see that the above bilinear terms of the scalar fields, in the frame that the four dimensional gravity action is canonical, are the kinetic terms of three complex scalar fields $ a_i, \,\, i \,=\, 0, 1, 2 \,\,$ belonging to three chiral multiplets which  stem from the  K\"{a}hler function
\bea
\Lambda \,=\, -2 \, ln (a_0 + a_0^*) \,- \, \frac{1}{2} \, ln (a_1 + a_1^*) \,- \, \frac{1}{2} \, ln (a_2 + a_2^*)
\label{kalit}
\eea
The complex fields $ a_i \,$ are expressed in terms of the $\,  \varphi, \sigma, \tau, A_5^1 , A_5^2$ and $\, A_5 \, $ fields by 
{\footnote{
These expressions are reminiscent of the corresponding used in \cite{Falkowski1} where $D=6$ supergravity, unlike in our considerations,  is compactified directly to a four dimensional supergravity 
}}
\bea
&& a_0 \,= \, \frac{1}{\sqrt{2}} \left( e^{\tau -\sigma /3} \,+\,  \frac{i}{2} \,  A_5  \right) \quad , \quad  
a_1 \,= \, \frac{1}{\sqrt{2}} \left( e^{\tau -\varphi + 2\sigma /3} \,+\,  \frac{2 \,i}{\sqrt{3}} \, A_5^1  \right) \; ,  \nonumber \\
&& \hspace*{2.3cm} a_2 \,= \, \frac{1}{\sqrt{2}} \left( e^{\tau +\varphi +2\sigma /3} \,+\, \frac{2 \,i}{\sqrt{3}} \, A_5^2 \right)
\eea

Note that in this case the "radion" field is expressed in a non-linear manner in terms of the fields $a_i$ as follows 
\bea
\tilde{e}_5^{\dot{5}} \,=\, \frac{1}{\sqrt{2}} (a_0 + a_0^*)^{2/3} (a_1 + a_1^*)^{1/6} (a_2 + a_2^*)^{1/6}
\label{radit}
\eea
Such a dependence of the radion field, in terms of the moduli fields, has been recently pointed out in \cite{Sakamura}. 
The K\"{a}hler function (\ref{kalit})  is of the no-scale type. In the absence of other superpotential couplings it would yield a vanishing scalar potential. It should be noted that the arising no-scale structure is due to the manner the four - dimensional supregravity is conceived as reduction of a higher supergravity theory. In our approach we first derived the $N=2, D=5$ supergravity  from a (1,0) six-dimensional supergravity by Kaluza-Klein compactification, keeping only the zero modes,  and then passed to a four-dimensional description on the branes  by orbifolding on $\, S_1/ Z_2$. Other schemes  \cite{Falkowski1, Para}, in which $D=6$ supergravity is compactified directly to a four dimensional supergravity,  do not lead to models bearing the aforementioned no-scale structure. The importance of having a no-scale structure is intimately connected with a positive scalar potential. 
In  schemes where the connection to the six-dimensional supergravity is absent, the $N=2, D=5$ supergravity, when compactified on $\, S_1/ Z_2$, has a  radion multiplet whose scalar component has a real part  linearly related to $\, \tilde{e}_5^{\dot{5}} \, $, see (\ref{radion}).

The couplings of the chiral multiplets living on the branes, of the model under study, is then obtained by considering the K\"{a}hler function
\be
\mathcal{K} \,=\, \Lambda + \frac{\delta (x_5)}{\tilde{e}_5^{\dot{5}}} \, K_V(\varphi _V, \varphi _V^*) + \frac{\delta (x_5 - \pi R)}{\tilde{e}_5^{\dot{5}}} \, K_H(\varphi _H, \varphi _H^*)
\label{kahler}
\ee
where $\varphi _V$ and $\varphi _H$ denote collectively the scalar fields of the chiral multiplets introduced on the visible and the hidden branes respectively.
In order to complete the brane-bulk Lagrangian we have to consider the superpotential terms on the branes as well. As explained in \cite{peggy2}, introducing a superpotential $W$ on any of the branes the potential induced is found from the expression
\bea
\Delta _{(5)}^2 e^{\mathcal{K}} \left\{ \mathcal{K}^{\alpha \beta ^*} \left[ W_ \alpha + \mathcal{K}_ \alpha  W \right] \left[ \bar{W}_{\beta ^*} + \mathcal{K}_{\beta ^*} \bar{W} \right] - 3\; W \bar{W} \right\} 
\label{poten}
\eea
In this $\Delta _{(5)}$ are the relevant scalar delta function defined as usual by $\Delta _{(5)} = \delta (x_5)/\tilde{e}_5^{\dot{5}} $, or $\Delta _{(5)} = \delta (x_5 - \pi R)/\tilde{e}_5^{\dot{5}} $,  depending on the brane, and the indices $\alpha , \beta $ indicate $a_i$ or brane fields.
A straightforward computation shows that the negative term in the potential given by Eq.(\ref{poten}) is canceled out leaving a scalar potential which is positive definite. 
In particular the inverse K\"{a}hler matrix $\, \mathcal{K}^{ \beta ^* \alpha} $, evaluated as an expansion in powers of $\Delta _{(5)}$ satisfying
\be
\mathcal{K}_{\alpha \beta ^*} \, \mathcal{K}^{ \beta ^* \gamma } \, = \, \delta _\alpha ^\gamma  \quad ,
\ee
when  plugged into (\ref{poten}) above, leads to elimination of the negative term  and a positive scalar potential 
{\footnote{
The ellipses denote terms of higher order in $\Delta _{(5)}$.
}}
\be
V_{scalar} \, = \, \Delta_{(5)} \, e^{\mathcal{K}} K^{m n*} W_m \bar{W}_n \, \left( \, 1 \, + \, \frac{1}{3} \, \Delta _{(5)} \, K^{m n*} K_{m} {K}_{n^*} \, + \, \cdots \right)
\label{scalarpot}
\ee
emerges. In it  the indices $m, n$ run now over the brane fields only.
The details of the calculation that leads to this form for the scalar potential is presented in the Appendix.

We see that as far as the potential and consequently the process of supersymmetry breaking is concerned, we have the same situation as in the case of the  N=2, D=5 supergravity discussed in the beginning which stands alone without reference to a higher dimensional supergravity it has its origin in. The reason behind this  result is due to the fact that the "K\"{a}hler"   function $\Lambda $ of Eq. 
(\ref{kalit}), which describes the Lagrangian of the restriction of the three complex fields on the branes, can be also written, using (\ref{radit}), as 
\be
\Lambda \,=\, - 3 \, ln \,(\tilde{e}_5^{\dot{5}}) \quad .
\ee
This states that $\, \Lambda$ is expressed in terms of the radion field in the same way as in previous considerations,  however now the radion is a non-trivial function of the fields $a_i$.

The above result can be generalized in the case of an arbitrary number $n_T$ of tensor multiplets. In this case the indices of the metric $G_{rs}$ in (\ref{sixdl}) run from one to $n_T + 1$ and  
the terms involving the field strengths $\, \hat{H}^{s \,\hat{M} \hat{N} \hat{P}} $ ( compare with (\ref{sixdl}) ) are again given by 
\bea
\hat{e}^{-1} \mathcal{L}^{(6)} \,&=&\,  \,-\, 
\frac{1}{12} G_{rs} \hat{H}^{r}_{\hat{M} \hat{N} \hat{P}} \hat{H}^{s \,\hat{M} \hat{N} \hat{P}} + ...  \nonumber 
\label{sixdl2}
\eea
Then a diagonalization of the metric $\, G_{rs} $  is necessary in order to bring these terms to a diagonal form as was done for the case of one tensor multiplet $\, n_T=1$.  
That done Eq. (\ref{fivedl}) is generalized to 
\begin{eqnarray}
\mathcal{L}^{(5)}_{0b} \,&=&\, \sqrt{-\tilde{G}} \,\, \bigg[ \frac{1}{2} \, R^{(5)}(\tilde{G}) - \frac{1}{3} \, \tilde{G}^{MN} \partial _M \sigma  \partial _N \sigma 
- \frac{1}{4} \, \tilde{G}^{MN} \partial _M \varphi  \partial _N \varphi \nonumber \\
&-& \frac{1}{16} \, e^{2\sigma /3} \, F_{MN} F^{MN} - \frac{1}{12} \, e^{ - 4\sigma /3} \, \sum_{r=1}^{n_T+1} \, f_r(\varphi_i)  \, F^{(r)}_{MN} F^{(r) \,MN}
 \bigg]
\label{fivedl2}
\end{eqnarray}
where in the base of a diagonal metric  the elements  $f_r(\varphi_i) $, which are positive definite,  depend on the scalar fields of the tensor multiplets $\, \varphi_i \,$ under consideration.  
In this case we have $n_T + 2$ chiral multiplets living on the branes, including the one stemming from the metric, and their corresponding complex scalar fields are given by
\bea
a_0 \,= \, \frac{1}{\sqrt{2}} \left( e^{\tau -\sigma /3} \,+\, \frac{i}{2} A_5  \right), \,\,
a_r \,= \, \frac{1}{\sqrt{2}} \left( \frac{1}{\sqrt{f_r}}e^{\tau  + 2\sigma /3} \,+\, \frac{2 \, i}{\sqrt{3}} A_5^r  \right).
\eea
In this case the "radion" field is expressed as a combination of all the fields $a_i$ involved as
\bea
\tilde{e}_5^{\dot{5}} \,=\, \frac{1}{\sqrt{2}} (a_0 + a_0^*)^{2/3} \prod _{r = 1}^{n_T + 1}(a_r + a_r^*)^{\frac{1}{3(n_T+1)}} \quad 
\label{modus}
\eea
This generalizes the previous result (\ref{radit}) for the case of one tensor multiplet. 
This is derived quite easily by first  eliminating $\sigma$, which is done  by combining the  real parts,  and using the fact that  $\prod f_r = 1$ since the metric determining the kinetic terms of the  tensor fields  and the accompanying scalars has unit determinant. Now the  "K\"{a}hler"   function $\Lambda $, which describes the restriction of the $n_T + 2$ complex fields on the branes, is  given by
\be
\Lambda \,=\, -2 ln (a_0 + a_0^*) \,- \, \frac{1}{n_T+1}\sum_{r=1}^{n_t+1} \, ln (a_r + a_r^*)  \,=\, -3 ln(\tilde{e}_5^{\dot{5}}).
\ee
From this form the no-scale structure  of the five-dimensional moduli, in the effective four-dimensional theory, is already recognized \cite{ovrut, Sakamura}. The novel feature in this consideration, which has its root in the the six-dimensional origin of the model at hand,   is that the effect of all  moduli fields, stemming from the tensor multiplets, are combined in a particular way to be expressed in terms of the "radion" component  $\,\tilde{e}_5^{\dot{5}} $ in the way displayed by  eq. (\ref{modus}). This result  could not have been anticipated. 

\section{The mediation of the supersymmetry breaking}
From the discussion presented in previous chapters, concerning the mechanism of the sypersymmetry breaking, it is apparent that this  
can be transmitted to the observable fields on the visible brane via gravitational interactions through loops including the gravitino fields as well as fermions of other chiral multiplets as already outlined in section 1. 
Unlike the $N=2$ $D=5$ case, where only  the radion and the gravity multiplet convey the message of the supersymmetry breaking, now due to the underlying structure of the higher dimensional supergravity  we have more messengers participating, for each multiplet accommodating the scalars $a_i$, but the qualitative features remain essentially the same. In the absence of supersymmetry breaking at the hidden brane, the gravitino is massless and no supersymmetry breaking terms are induced for the observable fields since  bosonic and fermionic loops cancel each other due to supersymmetry.  
This states that only fermion loops need be considered, as will be explained later, in which the gravitino and other fermionic states, like the fermion of the radion multiplet, which acquire masses via the superHiggs effect, are exchanged. 

The fermionic components can be grouped as four component symplectic Majorana Dirac spinors in the following way
\be
\Psi_m = \left( \begin{array}{c} \psi _m ^{1 }  \\\bar{ \psi} _{m}^{2 }  \end{array} \right) \quad , \quad 
 \Psi_j = \left( \begin{array}{c} \chi_j  \\\bar{ \xi}_j \end{array} \right) \, \, , \quad j = 1, \, ...  \, n_T+1
 \label{spinors}
\ee
where $\psi _m ^{1, \, 2}$ are the gravitinos with the flat indices $m$ taking values $\, 0,1,2,3$. The rest $\, \chi_j \, , \, \xi_j $  include proper linear combinations of the  fermions belonging to the chiral multiplets 
which have $\, a_j$  as their scalar components. In the absence of any tensor multiplet, there are only two such fermionic states which are the fifth components of the two five-dimensional gravitinos, namely $\, { \psi}_{\dot{5}}^{1} \,$, which is $ Z_2 - \text{odd} \,$, and the $  \, { \psi}_{\dot{5}}^{2} \,$ which is $ Z_2 - \text{even} $,   forming a  one Dirac state $\,\Psi_1$. In  writing  
Eq. (\ref{spinors}) we follow the convention where the  upper components  consist of even fields. 

For the calculation of the induced soft masses, the pertinent interaction  terms on the visible brane  are
\be
 - i \, k_5^2  \, e^{(4)} \,  K_V( \varphi,  \varphi ^* ) \, 
  \left[ \, 
 \bar{\Psi}_m  \, \Gamma^{mn} \,  \slash{\partial } \, \Psi _n \, + \,  \bar{\Psi}_m \, \Gamma^{m , \, j} \, \slash{\partial } \, \Psi_j \, + \, 
 \bar{\Psi}_i \, \Gamma^{ij}  \, \slash{\partial } \, \Psi_j
  - h.c 
\right] \; ,
 \label{ffk}
\ee   
while for the calculation of the induced soft trilinear couplings the relevant terms are  
\be
 -  \, k_5^2  \, e^{(4)} \,  W_V( \varphi ) \, 
  \left[ \, 
 \bar{\Psi}_m  \, \Sigma^{mn}  \, \Psi _n \, + \,  \bar{\Psi}_m \, \Sigma^{m , \, j} \,  \Psi_j \, + \, 
 \bar{\Psi}_i \, \Sigma^{ij}  \,  \Psi_j
  - h.c 
\right] \; ,
 \label{ffk2}
\ee   
{\footnote{
Eq. (\ref{ffk2}) holds for the hidden brane, as well, with $\, W \,$ replaced by the hidden brane constant superpotential $\, W_H = c \, $
}}
The subscript $V$ in $\, K_V, \, W_V \,$ denotes "visible" brane quantities. Note that there are no terms like (\ref{ffk}) on the hidden brane since all observable fields are confined  on the visible brane.  Terms of the form (\ref{ffk2}) do exist on the hidden brane, with $\, W_V$ replaced by the hidden brane constant superpotential $\, W_H = c \,$, which gives rise to fermion mass terms. However no fermion mass terms exist on the visible brane
{\footnote{
All scalars $\varphi$ have vanishing vev's, even if  Higgs fields are included, since there are no negative mass squared at this stage to drive electroweak symmetry breaking. Therefore 
$\, W_V(\vev{\varphi}) = 0 \,$.
}}
.

Before proceeding any further we should clarify our approach in order to find the loop contributions that are 
relevant for the mediation of the supersymmetry breaking. The only supersymmetry breaking terms in the Lagrangian are bilinear  fermionic mass terms (\ref{ffk2})  residing on the hidden brane, sourced by the constant superpotential 
$\, W_H = c \, $, which sets the order parameter of the supersymmetry breaking. Equivalently one can use the gravitino mass, instead of the parameter $\, c$ because they are proportional to one other \cite{bagger1}. It is evident from this that the brane to brane mediation of the supersymmetry breaking is implemented by fermion loops, since only fermions can carry the message of the supersymmetry breaking through their couplings with $\, W_H \, $. In order to segregate the part of the fermionic contribution that breaks supersymmetry from the rest in what follows we shall treat  the fermions as massless particles and their mass terms, which are proportional to the gravitino mass and are the only sources of the supersymmetry breaking, as vertices. In this way only fermionic loops involving  these vertices as  mass insertions yield contributions that break supersymmetry. 
Note that the mass insertions are located only on the hidden brane where  supersymmetry breaking takes place. 
Other fermionic contributions do exist but they are supersymmetric in nature, not depending on the gravitino mass, 
$\, m_{3/2}$, and hence not vanishing in the limit of  $\, m_{3/2}$ going to zero. These latter contributions need not be calculated since they cancel against the corresponding bosonic loop contributions due to supersymmetry ( see for instance \cite{rattazzi} ). Therefore the procedure we employ in the following picks up the supersymmetry breaking contributions which is our principal goal.  

In (\ref{ffk}) we tacitly assume that the K\"ahler function  $\,  K_V( \varphi,  \varphi ^* ) $  of the observable fields can be expanded as  $\, K_V( \varphi,  \varphi ^* ) =  \varphi  \varphi^* + ... $ so that the first terms yield canonical kinetic terms for the scalar fields $\, \varphi \, $.  With $\,  k_5^2 \, $ appearing in front of the interactions above  the dimensionality of the fields 
$\, \Psi_m \, , \, \Psi_i \, $   is two,  that is their mass dimension in the five-dimensional space-time, which implies   their propagators are massless
 {\footnote{
In (\ref{ffk}) and (\ref{ffk2}) we have reinstated dimensions by introducing the five-dimensional gravitational constant $\, k_5^2$. The latter is related to the four-dimensional 
gravitational constant by $\, k_5^2 = V_5 \, k_4^2  $, where $\, V_5$ is the volume of $\, S_1/Z_2$. Recall that  $\,  k_4^2 \equiv 8 \pi / M_{Planck}^{\, 2} = 8 \, \pi \, G_N \, $ with 
$M_{Planck}  = 1.22 \times 10^{\, 19} \; GeV \,$.
}} 
.
In (\ref{ffk}, \ref{ffk2}) the matrices 
$\, \Gamma  $ and $\, \Sigma $ are combinations of  gamma matrices and $\, L, R $ projection operators in general and depending on the case they carry  indices $\,  j = 1, \, ... \, n_T+1 \, $  labeling the fermions $\, \Psi_j $, see  Eq. (\ref{spinors}). 
 
Since we are interested in fermion loops  connecting  the hidden to the visible brane, we need the  $\, \Psi_m \, , \, \Psi_j$ propagators. These fermionic states are mixed in the kinetic and / or mass terms, however by a suitable gauge choice the $\, \Psi_m \,$ disentangles from $  \, \Psi_j$ in the kinetic terms. Therefore we can calculate the gravitino and $\, \Psi_j $'s  massless propagators which are easy to find and treat the mass-mixing terms as insertions. That was also the treatment followed in  \cite{peggy2, peggy3}. Then we use the Dirac and gravitino propagators in the mixed momentum-configuration space representation \cite{arkani1, puchwein, meissner} which, suppressing  indices, have the generic form $\, \Delta(p, y, y^\prime ) \,  $, to be given below. In it $\, y, y^\prime $ label points along the fifth direction and $\, p$ is the four dimensional momentum. 
For the case of interest only propagations from one brane to the other and propagations on the same  brane are encountered so  $\, y, y^\prime $ can take values $\, 0 \, $ or $ \, \pi R$.  
In this case the massless propagators $\, \Delta(p, y, y^\prime ) \,  $ get  a rather simple form 
\be
G_{mn}(p,y,y') = \lambda \, \, i \, \gamma _n \, \slash{p} \, \gamma _m \, \, S(p,y,y') \quad , \quad 
G_{ij}(p,y,y') = \lambda_{ij} \, \, i \, \slash{p} \, \, S(p,y,y')
\label{propa}
\ee
In these $\, \lambda , \lambda_{ij} $ are dimensionless constants  and the function $\, S(p,y,y') $ is given by 
\be
S(p,y,y') \equiv  \frac{cos \, q (\pi R - \mid y - y' \mid )}{2 \,q \,sin(q \pi R)} \quad , \quad q \equiv \sqrt{ - \, p^2 }
\label{propa2}
\ee 
In (\ref{propa}) $\, G_{mn}$ and $\,G_{ij}$ are the the propagators of the gravitino and the fermions $\, \Psi_i$. Note that these carry zero dimensionality in conform with the fact that 
the fermions carry dimension two, as already discussed. 
\begin{figure}  
\begin{center}
\includegraphics[scale=0.5]{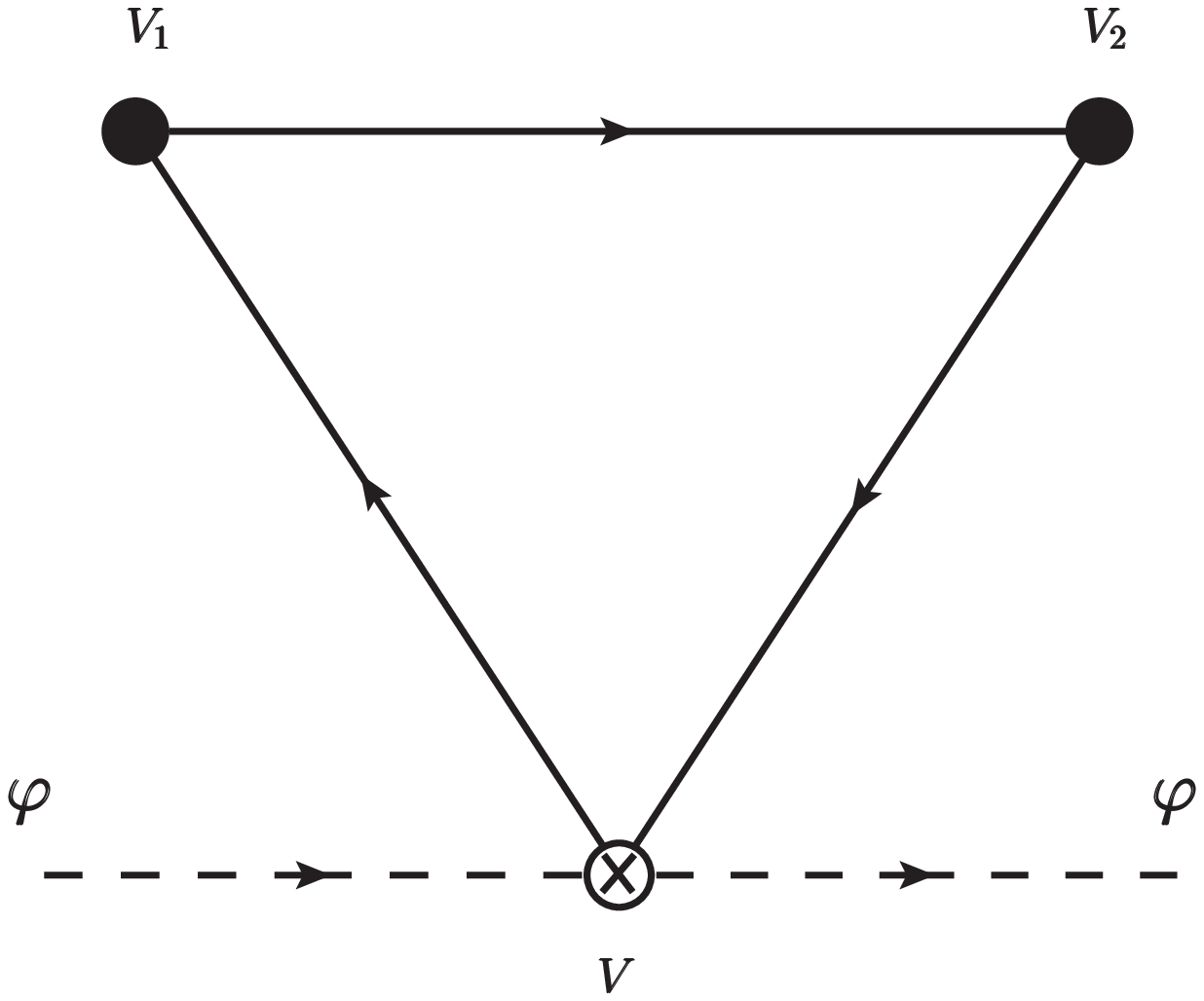}
\hspace*{2cm}
\includegraphics[scale=0.5]{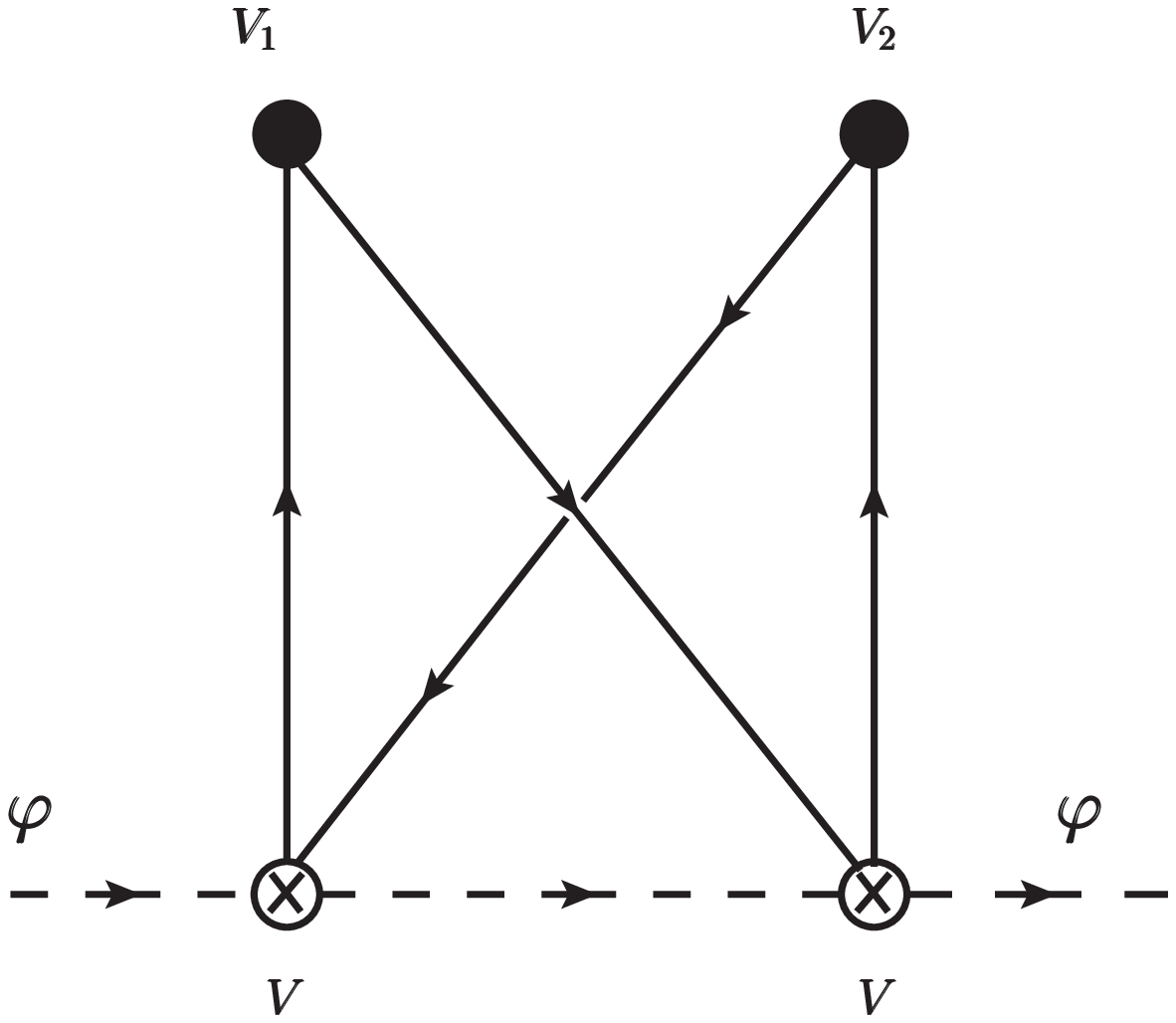}
\end{center}
\caption[]{
The mechanism of mediation of supersymmetry breaking from the hidden brane ( solid vertices on top ) to the visible brane ( crossed vertices at bottom  ) via exchanges of fermions and gravitinos  which are represented by solid lines. The dashed lines represent scalar particles living on the visible brane. A one-loop graph ( left pane ) and a two-loop graph ( right pane ), giving rise to a scalar mass, are displayed. 
}
\label{figurex}  
\end{figure}

In figure \ref{figurex} we give  examples of mass corrections to the visible scalar fields. In this, lower vertices $\, V$ are on the visible brane and upper vertices $\, V_{1,2} \, $ are located on the hidden brane. The dashed lines represent scalar fields on the visible brane and the solid lines all possible fermion propagators that can be drawn. The vertices $\, V$, stemming from the interaction (\ref{ffk}), give rise to soft scalar masses squared. Similar graphs with $\, V$ replaced by the interaction (\ref{ffk2}) induce trilinear soft couplings. Allowing for gauge fields on the brane gaugino masses can arise in a similar manner.   
In that case gaugino fields can be  also exchanged ( see \cite{peggy3}). From the one-loop graph on the left pane of \ref{figurex} we have a contribution which, suppressing all indices, for simplicity,  is given by
\be
m_0^2 \, = \, \int \, d^4 p \, Tr \, \left[ \, V \, \Delta(p, 0, \pi R) \, V_1  \, \Delta(p, \pi R , \pi R) \, V_2 \,  \Delta(p, \pi R , 0) \, \right]
\label{struct}
\ee
External momenta of the scalar fields have been put to zero. The graph on the right pane has a similar structure. In (\ref{struct}) $\, \Delta$ denotes either a gravitino or a fermion propagators in all possible ways allowed by the interactions, and the mass terms  are treated as insertions. For the graphs displayed in \ref{figurex} the vertices are $\, V = k_5^2 \, \slash{p}  $ and 
$\, V_1 = V_2 = k_5^2 \, c \,$ as follows from (\ref{ffk}) and  (\ref{ffk2}).
Thus up to a dimensionless constant (\ref{struct}) yields 
\bea
m_0^2 \, &=& \,i \, k_5^6 \, {| c |}^2 \, \int \, \dfrac{d^4 \, p}{ {(2 \pi)}^4} \, \dfrac{ p^4 \, cos ( q \pi R) }{ {( q \, sin ( q \pi R)  )}^3  } 
\, = \, k_5^6 \, {| c |}^2 \,  \dfrac{2 \pi^2}{{(2 \pi)}^4}   \, \int_0^\infty \, {d \, p_E \, p_E^3} \, \dfrac{ p_E^4 \, cosh ( p_E \pi R) }{ {( p_E \, sinh ( p_E \pi R)  )}^3  } 
\nonumber \\
 &=&  k_5^6 \, {| c |}^2 \, \dfrac{1}{ {(\pi R)}^5 \,  {(2 \pi)}^4  }    \, \int_0^\infty \, {d \, \xi} \, \dfrac{ \xi^4 \, cosh \xi  }{ { \, sinh \xi  }^3  } 
 \, = \, \dfrac{3 \, \zeta(3)}{8} \, g \, m_{3/2}^2
\eea
In  the first equation above $\, q \equiv \sqrt{-p^2} \,$ and in the second equation  we passed  to Euclidean momenta. Also the integration in the $\, \xi$ variable yields $\,  3 \, \zeta(3)$
 where  $\,   \zeta(3)$ is the Riemann zeta function. In passing to the the last expression, we used  the value of the lowest mode gravitino mass,   
$\, m_{3/2} \, = \, k_5^2 \, c / \pi R \, $,  and a  dimensionless coupling was defined by $\, g \equiv \dfrac{k_5^2}{\pi^2 \,V_5^3} \, $ with $\, V_5 \, $  the volume of the compactifying manifold  
$\, V_5 = \pi \, R $. Then the soft mass squared above is of the form 
\be
m_0^2 \, = \, C_1 \, g \, {m}^2_{3/2}
\ee
where $\, C_1 \, $  is a finite non-zero dimensionless constant of order unity. The two loop graph on the right pane of \ref{figurex} can be handled in the same way. In this case we have an extra vertex 
$\, V$ an extra fermion propagator and a scalar propagator on the visible brane. Its contribution is 
\bea
 m_{0,(2-loop)}^2 \, = \hspace*{12cm} \nonumber \\
  \, k_5^8 \, {| c |}^2 \, \int \, \dfrac{d \, p_1^4}{ {(2 \pi)}^4} \, \int \, \dfrac{d \, p_2^4}{ {(2 \pi)}^4} \,
\dfrac{ p_1^{\, 2}   }{ {( q_1 \, sin ( q_1 \pi R)  )}^2  } \,  \dfrac{ p_2^{\, 2}  }{ {( q_2 \, sin ( q_2 \pi R)  )}^2  } \, \dfrac{{( p_1+ p_2 )}^{\, 2}} { {( p_1-p_2 )}^{\, 2}} 
\label{2loop}
\eea
The momenta $q_i$ above are $\,  q_i \equiv \sqrt{ - p_i^{ \, 2} } $, see Eq. (\ref{propa2}). Also $\, {( p_1+ p_2 )}^{\, 2}$ in the numerator comes from the two $V$-vertices while  
$\, {( p_1- p_2 )}^{\, 2}$ in the denominator from the boson propagator on the visible brane. 
Notice that there are no $\, cosine  $ terms in the numerator, as the ones appearing in  (\ref{propa2}), since all fermion propagators connect the visible to the hidden brane in this case. Certainly it is not our intention to proceed to a two loop calculation of the soft SUSY-breaking parameters and the only  reason of presenting (\ref{2loop}) is to derive an order of magnitude estimate that will be useful for the discussion that follows.  Proceeding in the same way as before it is not hard to see, by first passing to Euclidean momenta and then rescaling them, in order to have dimensionless integrals, that a correction arises  which is proportional to $\, \sim g^{\,2} \, {m}^2_{3/2} $. This argument is sufficient for one to be convinced that the dimensionless coupling $\, g $ defined before serves as a proper loop expansion parameter. This can be expressed in terms  of the Planck mass  as 
\be
g \, \equiv \, \dfrac{k_5^2}{\pi^2 \,V_5^3} \, = \, \dfrac{ 1 }{ \pi^3 \, R^{\,2} \, M_{Planck}^{\, 2} }
\label{ggiven}
\ee
Therefore from the previous discussion it is evident that the induced soft scalar masses can be written, in general, as 
\be
m_0^2 \, = \, (  \; \sum_L \, C_L \, g^L \; )  \;  {m}^2_{3/2}
\label{softm}
\ee
Similar results hold for the soft gaugino masses and the trilinear couplings as well, 
\be
M_{1/2} \, = \, (  \; \sum_L \, C_L \, g^L \; )  \;  {m}_{3/2} \quad , \quad
A_{0} \, = \, ( \; \sum_L \, C_L \, g^L \; )  \;  {m}_{3/2} 
\label{softam}
\ee
The constants $C_k$ are dimensionless and they are different, in general, for each of the parameters  $\,m_0^2 , \, M_{1/2} , \, A_0 \,  $.
{\footnote{
Although $\, C_1$ are finite at the one-loop order we do not know whether $C_k$'s are finite to higher loop orders.  
Therefore our arguments hold provided that this is true.
}}. 
It is obvious from these results that loop expansion is a valid approximation provided 
\be
g  < 1 
\label{bound1}
\ee
or, on account of (\ref{ggiven}),  when  $\, R^{-1} < \pi^{3/2} \, M_{Planck} \, $ which allows for large values of the radius  $\, R \,$.

It should be pointed out  that, the corrections to the soft parameters presented before are valid to any loop order but only to lowest order in the gravitino mass. However, there are corrections that are of higher order in the gravitino mass. For instance, for the gaugino mass, if the quadratic in the gravitino mass corrections are added, we have 
\be
M_{1/2} \, = \, C_1 \, g \,  m_{3/2} \, ( \, 1 \, + {\cal{O}}(1) \, \,R \, m_{3/2} \,) \, \; .
\ee
Therefore the calculation of the induced soft-supersymmetry breaking terms, to lowest order in the gravitino mass, can be trusted provided
\be
R \, m_{3/2} < 1
\label{boundx}
\ee
Thus  in this approximation, given the value of $\, R$, an upper  bound  is imposed on the gravitino mass
 {\footnote{
This bound is the consistent with  $m_{3/2} = k_5^2 c / \pi R$ which also  holds in this regime ( see footnote \ref{footot} ).
}}
. 
In the regime (\ref{boundx}),  corrections that are of higher power in the gravitino mass  are irrelevant. This facilitates calculations a great deal. 
Therefore we conclude that in addition to (\ref{bound1}), which guarantees a valid perturbative expansion, the bound given by (\ref{boundx}) should be observed so that the resulting soft breaking terms are trusted to lowest order approximation in the gravitino mass.  Thus the region of the parameters for which the scheme presented makes sense is defined by  (\ref{bound1}) and (\ref{boundx}) and these limits are important for  phenomenological analysis 

Using the bound (\ref{boundx}) it is found that  the induced gaugino mass $\, M_{1/2} \, \sim \, g  \, m_{3/2} $  is also  bounded from above by
\be
M_{1/2} \,  < \,  g \, R^{\,- 1} =  \,  \dfrac{R^{-3}}{\pi^3 \; M_{Planck}^2} 
\label{bound2}
\ee
while similar bounds hold for the other soft breaking parameters as well. The absence of any supersymmetry signals at LHC  indicates that the upper bound (\ref{bound2}) set for $\, M_{1/2}$ should be larger than a few  $\TeV$, at least. Therefore  (\ref{bound2}) sets a lower bound on  $\, R^{\, -1}$ which combined with the upper bound from (\ref{bound1}) sets the allowed range of values for the size of the orbifold, 
\be
\pi \; M_{1/2}^{1/3} \; M_{Planck}^{2/3} \, < \, R^{-1} \, < \, \pi^{\,3/2} \, M_{Planck} 
\label{bound3}
\ee
From this we derive a  lower limit  $\, R^{\, -1} \, \gtrsim 1.66 \times  10^{\,11} \, $ TeV if $\,  M_{1/2} \sim 1 \, \text{TeV}$ . 
Therefore  the radius $\,R$ can be large but not larger than $\, \sim \; 10^{\, - 11} \, {\mathrm{TeV}}^{-1} $, if  supersymmetric masses  are heavier than $\sim$ TeV. For a given value of $\, R$ in this regime the value of $\, g$ is determined by (\ref{ggiven}). 

\begin{table}
\addtolength{\tabcolsep}{2pt}
\begin{tabular}{c|c|c|c|c} 
\hline
{\rule{0pt}{4.4ex}}
 $\quad \quad \quad R^{\, -1}  \quad \quad $   &  $ \, \quad \quad g  \quad \quad $       &   $\, m_{3/2} $  &  $ M_{1/2} , \, \, A_{\,0} \sim g \, m_{3/2} $   & $ m_{\,0} \sim \sqrt{g } \; m_{3/2} $  
\\ [14pt] \hline \hline 
{\rule{0pt}{4.5ex}}
  $10^{\, 11} $   \text{  TeV}    &     $10^{\, - 10} $    &      $ 10^{\, 10} \, - \, 10^{\, 11} $    &   $ 1 \, - \, 10 $     &   $ 10^{\, 5} \, - \, 10^{\, 6} $   \\ [2pt]
                             &          &         &           &     \\ [2pt]\hline
{\rule{0pt}{4.5ex}}
 $10^{\, 14} $   \text{  TeV}    &       $10^{\, - 4} $  &      $ 10^{\, 4} \, - \,  10^{\, 14} $      &   $ 1 \, - \, 10^{\,10} $    &   $ 10^{\, 2} \, - \, 10^{\, 12} $   \\ [2pt]
                            &          &         &             &     \\ [2pt]\hline
{\rule{0pt}{4.5ex}}
 $10^{\, 16} $   \text{  TeV}   &        $ 1 $   &     $ 1 \, - \,  10^{\, 16} $     &    $ 1 \, - \, 10^{\,16} $      &  $ 1 \, - \, 10^{\,16} $   \\
                             &          &         &          &   \\ [2pt]\hline
\end{tabular}
\caption{
{\normalsize{
In the first column the value of $\, R^{\, -1}$ is given ( in TeV ). In the second column the corresponding value of the loop expansion parameter $g$ is given and in the third the allowed range for the gravitino mass is shown. In the the fourth and fifth columns the corresponding allowed ranges for the soft SUSY-breaking parameters are given. All masses are in TeV .
}}
}
\label{tablecharter}
\end{table}

The soft SUSY breaking parameters, and consequently the mass spectrum of supersymmetric particles, is determined when in addition to $R$  the value of the gravitino mass is given, which however should be smaller than $\, R^{\, -1}$ by (\ref{boundx}). In Table 
\ref{tablecharter} we display representative values of $\, R$, the corresponding values of $\,g$ and the allowed ranges for the gravitino mass and the soft breaking parameters. The lowest value of the gravitino mass is dictated by the fact that all supersymmetry breaking parameters are larger  than one TeV. For the lowest allowed value of  $\, R^{\, -1}$, displayed in this table, the parameter $\, g$ is tiny and the gravitino mass is constrained within rather narrow limits. Then the value of the gaugino mass is in the range of a  few TeV.  Note however that the soft mass $\, m_0$ is much larger. The reason  is that the soft SUSY-breaking parameters have a different dependence on  $\, g$. In particular $\, M_{1/2} $ and $\, A_{0} $ are linear in $\, g$ while $\, m_0$ is proportional to $\, \sqrt{g}$ and the soft masses are correlated having a ratio $\, M_{1/2} / m_0 \sim \sqrt{g} \; $, which is independent of the gravitino mass. Thus  smaller values of $g$ entail to larger masses for  $\, m_0$ as compared to the gaugino masses. This forces $\, m_0$ to be considerably larger than  $\, M_{1/2} $,  for values of  $\, R^{\, -1}$ in the range up to $\, 10^{\, 14} $ TeV. For higher $\, R^{\, -1}$ the soft mass $\, m_0$  is still larger than the other SUSY breaking parameters but not much. This is important for phenomenological studies since it shows that for small values of  $\, R^{\, -1} $ the squarks and sleptons can be significantly heavier than the gauginos. This mimics the models of split supersymmetry \cite{Giudice:2004tc,Wells:2003tf,split,Antoniadis:2004dt} and shows that in the proposed six-dimensional supergravity framework  a split supersymmetry  scenario can be accommodated,  in principle. 
 
As is seen from the last row of Table \ref{tablecharter}, as the value of $\, R^{\,-1}$ approaches the Planck scale the restrictions put on the gravitino mass, other than the phenomenological ones, are loosen and $\, m_{3/2}$ can take any value from the TeV to the Planck scale. In this case $g \simeq 1$ and  the effective  supersymmetry breaking scale in the observable sector is of the same order with the gravitino mass. 

\section{Conclusions}

In this note, motivated by the recent developments pertinent to the six-dimensional description of the supergravity theory, we study the orbifolded $N=2$, $D=5$ supergravity  as originating from a $D=6$ supergravity. In particular 
we considered a six-dimensional supergravity model, with tensor multiplets, which is truncated  to a five-dimensional supergravity  by the Kaluza - Klein  approach, followed by orbifold compactification on $\, S_1/Z_2$. The arising model is described by a five-dimensional supergravity  with two four-dimensional branes residing at the end points of $S_1/Z_2$, having local $N=1$ supersymmetry on them. 
The resulting model has striking similarities with  supergravitiy models studied in the past  that they are not connected to higher dimensional schemes.  The novel features that  the connection with the six-dimensional supergravity brings about, is the fact that additional chiral multiplets appear which can convey  the messenge of supersymmetry breaking from one brane to the other. Besides, the particular compactification approach results to an effective supergravity on the four-dimensional branes with special characteristics. In particular the additional scalar fields, whose  presence in due the tensor multiplets, parametrize a K\"{a}hler manifold having a no-scale structure which could not have been anticipated. The no-scale structure is dictated by the radion field, which is a combination of the moduli fields, and it is  derived from the six-dimensional theory. 

The supersymmetry breaking that takes place on the hidden brane is transmitted through the bulk by gravity to the observable fields which are assumed to be confined on one of the branes. The magnitude  of the 
supersymmetry breaking is set by the gravitino mass and the size $R$ of the orbifold $\, S_1/Z_2$. The mechanism of supersymmetry breaking transmission reveals that finite soft supersymmetry breaking parameters are induced for the supersymmetric fields living on the visible brane. The perturbative  mechanism of supersymmetry breaking can be trusted at one - loop approximation and to lowest order in the gravitino mass for values of $R^{\, - 1}$ within a certain range which can be as large as the Planck scale but not smaller than $\sim 10^{\,11} \TeV $, if sparticles have to have masses larger than a few TeV as LHC experiments suggest. The gravitino mass is forced to lie within a narrow  mass range, in the vicinity of $\, \sim 10^{\, 10} \, \TeV $, for the lowest allowed value of  $R^{\, -1} \sim 10^{\,11} \TeV $, but this range is  broadened if $\, R^{\, -1}$ gets larger. 
The above scheme allows for  sparticles  to be considerably heavier than sfermions, when $R^{\, - 1} \lesssim 10^{\,14} \TeV $, mimicking the models of split supersymmetry. For values of $R^{\, - 1}$ approaching the Planck scale the gravitino mass and the corresponding SUSY breaking parameters  are of the same order of magnitude and  can take values in the whole range from a few TeV  to $10^{\, 16}$ TeV. 

The salient qualitative features of the five-dimensional schemes regarding the supersymmetry breaking remain almost intact, indicating that  the supersymmetry breaking mechanisms studied in the past can be easily incorporated in higher dimensional schemes. A more detailed theoretical and phenomenological study of these models will appear in a forthcoming publication. 

\vspace*{.3cm}

\section*{Acknowledgments}

This research has been co-financed by the European Union (European Social Fund – ESF) and Greek national funds through the Operational Program "Education and Lifelong Learning" of the National Strategic Reference Framework (NSRF) - Research Funding Program: {\bf{THALES}}. Investing in knowledge society through the European Social Fund. \\
One of us A.B.L. thanks I. Antoniadis for illuminating discussions.

\newpage
\renewcommand\appendix{\par
\setcounter{section}{0}%
\setcounter{subsection}{0}%
\setcounter{table}{0}
\setcounter{figure}{0}
\setcounter{equation}{0}
\gdef\thetable{\Alph{table}}
\gdef\thefigure{\Alph{figure}}
\gdef\theequation{\Alph{section}.\arabic{equation}}
\section*{Appendix}
\gdef\thesection{\Alph{section}}
\setcounter{section}{0}}

\begin{appendix}
\section{The scalar potential on the brane}

For the calculation of the scalar potential given in (\ref{poten}) we need $\, \mathcal{K}^{\alpha \beta^*} $ and $\,\mathcal{K}_{\alpha ^*}  $, as well as its complex conjugate $\,\mathcal{K}_{\alpha }  $. The indices take values  $i=0,1,2$, corresponding to the fields $a_0, a_1, a_2$, and $\, m$ corresponding to a scalar $\phi_m$ living on the branes. 
For $\,\mathcal{K}_{\alpha }  $ we find from (\ref{kahler}) 
\bea
\mathcal{K}_{i} \, \; &=& \, - n_i \,( a_i + a_i^* )^{-1} \, ( \, 1 + \frac{1}{3} \, \Delta _{(5)} \, K \, ) \nonumber \\
\mathcal{K}_{m} \, &=& \, \Delta _{(5)} \, {K}_{m}
\label{qq1}
\eea
In these, $i =0,1,2$ and $\, n_0 = 2, n_1 = n_2 = 1/2 \,  $ and  $K$ is either $K_V$ or $K_H$ depending on which brane, visible or hidden, the scalar potential refers to ( see equation (\ref{kahler})). In the following we shall present the calculation for the scalar potential on the visible brane. The result can be applied for the hidden brane, as well, replacing the visible K\"{a}hler function and superpotential by the corresponding ones for the hidden brane. For definiteness in the following we shall assume that only one scalar field lives on the brane, $\varphi$.  This suffices to demonstrate how one arrives at the result presented in (\ref{scalarpot}). The elements of the inverse matrix  
$\, \mathcal{K}^{\alpha \beta^*} $ are given by
\bea
\mathcal{K}^{j i^*} \, &=& \, - n_i^{\, -1} \, \,( a_i + a_i^* )^{2 } \, \delta^{j i^*} \quad , \quad j, i^* = 0,1,2 \nonumber \\
\mathcal{K}^{\varphi i^*} \, &=& \,  ( a_i + a_i^* ) \, \frac{K_{\varphi^*}}{3 \,K_{\varphi \, \varphi^*}} \quad , \quad 
\mathcal{K}^{i \varphi^*} \, = \,  ( a_i + a_i^* ) \, \frac{ K_\varphi}{3 \, K_{\varphi \, \varphi^*}} \nonumber \\
\mathcal{K}^{\varphi \varphi^*} \, &=& \frac{1}{\Delta _{(5)}} \, \frac{1}{K_{\varphi \, \varphi^*}} \, + \,
\frac{K_{\varphi^*} \, K_\varphi}{3 \,{K_{\varphi \, \varphi^*}^2}}
\label{qq2}
\eea
In deriving (\ref{qq2}) we have omitted terms which are of order $\, \Delta_{(5)} $, or higher, the reason being that 
(\ref{poten}) already includes a prefactor $\, \Delta_{(5)}^2 $ and we are interested in potential terms up to second order in $\, \Delta _{(5)} $. 

 For the calculation of the scalar potential we first  write the expression in 
(\ref{poten}) which involves  $\, \mathcal{K}^{\alpha \beta^*} $ as
\bea
&&\mathcal{K}^{\alpha \beta ^*} \left[ W_ \alpha + \mathcal{K}_ \alpha  W \right] \left[ \bar{W}_{\beta ^*} + 
\mathcal{K}_{\beta ^*} \bar{W} \right] \nonumber \\
&&\quad \quad \quad \quad  
 \, = \, \mathcal{K}_{\alpha} \mathcal{K}^{\alpha \beta^*} \mathcal{K}_{\beta ^*} \,  {| W |}^2 \, + \, 
( \, \bar{W} \, W_\varphi \, \mathcal{K}^{\varphi \beta^*} \mathcal{K}_{\beta ^*} \, + \, c.c \, ) \, + \,
\mathcal{K}^{\varphi \varphi^*} \, W_\varphi \, W_{\varphi^*}
\label{qq3}
\eea
In writing this we have made use of the fact that the superpotential $W$ does not depend on the fields  $\, a_i$, therefore 
$\, W_i = \bar{W}_{i^*} = 0 $.
 Then having the analytic forms of the quantities given by (\ref{qq1}) and (\ref{qq2}) is fairly easy to see that 
\be
\mathcal{K}_{\alpha} \mathcal{K}^{\alpha \beta^*} \mathcal{K}_{\beta^*} \, = \, 3 \, + \, ... 
\label{rr1}
\ee
In it the ellipsis denote terms  of order $\, \Delta_{(5)} $,  which can be neglected if we keep terms in the potential that are most quadratic in $\, \Delta_{(5)} $. Recall that the potential, as given by equation (\ref{poten}), already carries a prefactor which is quadratic in $\, \Delta_{(5)} $. The result  (\ref{rr1}) states that the last negative term in the potential (\ref{poten}), namely  $\, - 3 \, {| W |}^2  $,  is exactly cancelled by the first term appearing on the r.h.s. of (\ref{qq3}). The terms in (\ref{qq3}) which involve $\, \mathcal{K}^{\varphi \beta^*} \mathcal{K}_{\beta ^*} $  are easily found to be proportional to  $\, \Delta_{(5)} $ and hence do not contribute to the desired order. Therefore it only remains to examine the last term $\, \mathcal{K}^{\varphi \varphi^*} \, W_\varphi \, W_{\varphi^*} $, which by using the expression for $ \, \mathcal{K}^{\varphi \varphi^*} $ in (\ref{qq2}), yields 
\bea
\mathcal{K}^{\varphi \varphi^*} \, W_\varphi \, W_{\varphi^*} \, &=& \, 
\left( \frac{1}{\Delta _{(5)}} \, \frac{1}{ K_{\varphi \, \varphi^*}} \, + \,
\frac{K_\varphi^* \, K_\varphi}{ 3 \,{K_{\varphi \, \varphi^*}^2}} \, \right) \, W_\varphi \, W_{\varphi^*} 
\nonumber \\
\, &=& \, 
\frac{1}{\Delta _{(5)}} \, {K^{\varphi \, \varphi^*}} \, W_\varphi \, W_{\varphi^*} \, 
\left( \, 1 + \frac{\Delta _{(5)}}{3} \, {K^{\varphi \, \varphi^*}} \mathcal{K}_{\varphi} \,\mathcal{K}_{\varphi^*} \right)
\label{endr}
\eea
In passing to the last equation we used the fact $\, {K^{\varphi \, \varphi^*}} \, {K_{\varphi \, \varphi^*}} =1 $. Using the above equation in (\ref{poten}) we arrive at 
\bea
V_{scalar} \, = \, \Delta _{(5)} \, e^{\mathcal{K}} K^{\varphi \varphi^*} W_\varphi \bar{W}_{\varphi^*} \, \left( \, 1 \, + \, \frac{1}{3} \, \Delta _{(5)} \, K^{\varphi \varphi^*} K_ \varphi \bar{K}_{\varphi^*} \, + \, \cdots \right)
\eea
which is just the result presented in equation (\ref{scalarpot}) for the case of one chiral multiplet. This result can be generalized for an arbitrary number of multiplets and it can be also used for the hidden brane as well, with $K$ and $W$ replaced by their corresponding quantities on the hidden brane. Since we have assumed a constant superpotential on the hidden brane  the  derivatives $\, W_\varphi \, , \, \bar{W}_{\varphi^*} $  are set to zero yielding a vanishing scalar potential on the hidden brane.  

\end{appendix}


\end{document}